\documentclass[sigconf]{acmart}
\AtBeginDocument{%
  }

\copyrightyear{2026}
\acmYear{2026}
\setcopyright{cc}
\setcctype{by}
\acmConference[CHI EA '26]{Extended Abstracts of the 2026 CHI Conference on Human Factors in Computing Systems}{April 13--17, 2026}{Barcelona, Spain}
\acmBooktitle{Extended Abstracts of the 2026 CHI Conference on Human Factors in Computing Systems (CHI EA '26), April 13--17, 2026, Barcelona, Spain}
\acmDOI{10.1145/3772363.3798643}
\acmISBN{979-8-4007-2281-3/2026/04}

\begin{document}

\title{Reading the Mood Behind Words: Integrating Prosody-Derived Emotional Context into Socially Responsive VR Agents} 

\author{SangYeop Jeong}
\authornote{Both authors contributed equally to this work.}
\orcid{0009-0004-0937-6490}
\affiliation{%
  \department{Department of Applied Artificial Intelligence}
  \institution{Seoul National University of Science and Technology}
  \city{Seoul}
  \country{Republic of Korea}
}
\email{yeobi5840@seoultech.ac.kr}

\author{Yeongseo Na}
\authornotemark[1]
\orcid{0009-0004-5516-902X}
\affiliation{%
  \department{Department of Applied Artificial Intelligence}
  \institution{Seoul National University of Science and Technology}
  \city{Seoul}
  \country{Republic of Korea}
}
\email{yeongseo0407@seoultech.ac.kr}

\author{Seung Gyu Jeong}
\orcid{0009-0007-2600-6389}
\affiliation{%
  \department{Department of Applied Artificial Intelligence}
  \institution{Seoul National University of Science and Technology}
  \city{Seoul}
  \country{Republic of Korea}
}
\email{wa975@naver.com}

\author{Jin-Woo Jeong}
\orcid{0000-0001-9313-6860}
\affiliation{%
  \department{Department of Data Science}
  \institution{Seoul National University of Science and Technology}
  \city{Seoul}
  \country{Republic of Korea}
}
\email{jinw.jeong@seoultech.ac.kr}

\author{Seong-Eun Kim}
\authornote{Corresponding author.}
\orcid{0000-0002-4518-4208}
\affiliation{%
  \department{Department of Applied Artificial Intelligence}
  \institution{Seoul National University of Science and Technology}
  \city{Seoul}
  \country{Republic of Korea}
}
\email{sekim@seoultech.ac.kr}

\renewcommand{\shortauthors}{Jeong et al.}

\begin{abstract}
In VR interactions with embodied conversational agents, users’ emotional intent is often conveyed more by how something is said than by what is said. However, most VR agent pipelines rely on speech-to-text processing, discarding prosodic cues and often producing emotionally incongruent responses despite correct semantics. We propose an emotion-context-aware VR interaction pipeline that treats vocal emotion as explicit dialogue context in an LLM-based conversational agent. A real-time speech emotion recognition model infers users’ emotional states from prosody, and the resulting emotion labels are injected into the agent’s dialogue context to shape response tone and style. We conducted a within-subjects VR study ($N=30$) using semantically neutral and emotionally ambiguous utterances, comparing an emotion-aware agent with a text-only baseline. Results show significant improvements in dialogue quality, naturalness, engagement, rapport, and human-likeness, with 93.3\% of participants preferring the emotion-aware agent. These findings suggest that prosody-derived emotional context enables socially responsive VR conversations.
\end{abstract}


\begin{teaserfigure}
  \includegraphics[width=\textwidth]{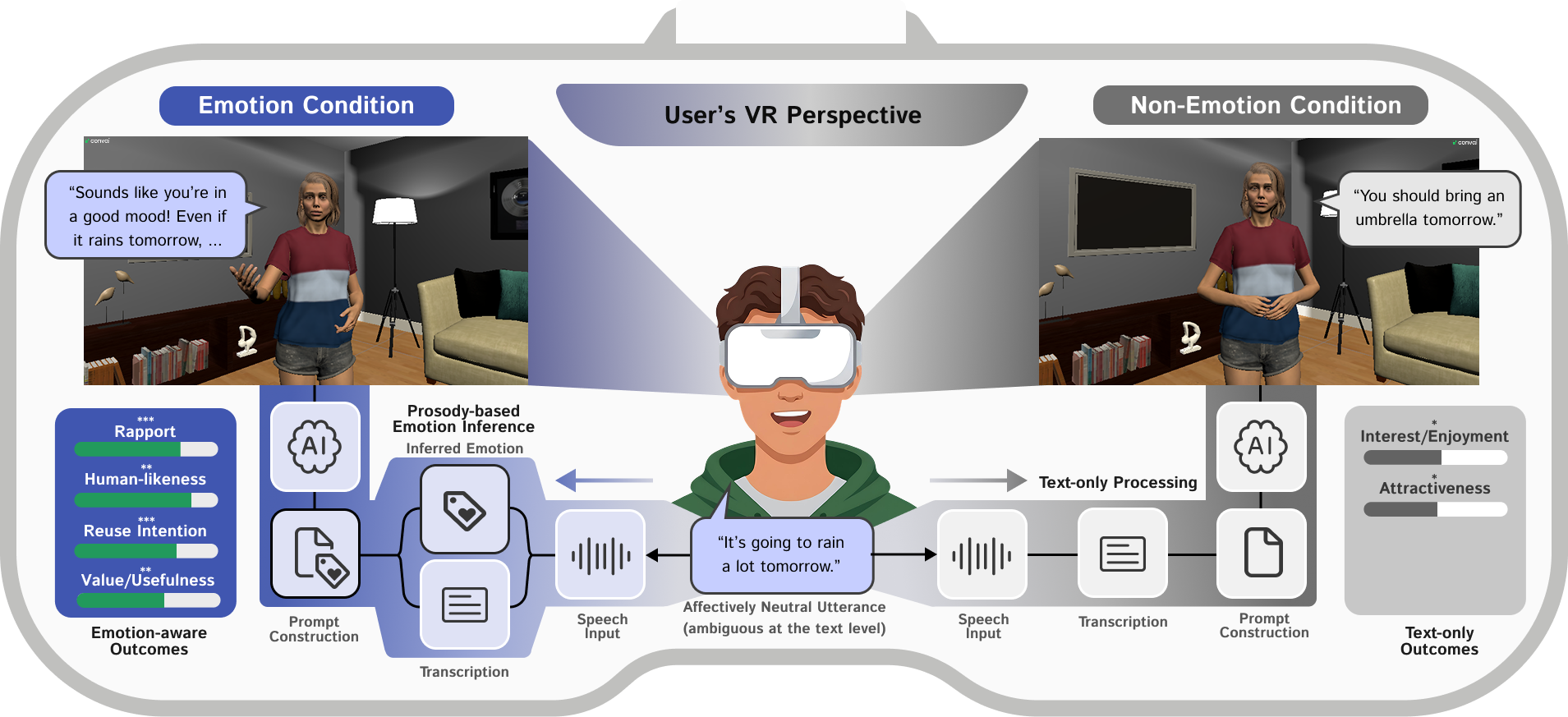}
  \caption{Overview of the emotion-aware interaction pipeline in VR: The Emotion Recognition (ER) condition (left) injects Speech Emotion Recognition (SER)-derived labels into the LLM prompt to ensure affective congruence, whereas the Non-Emotion Recognition (NER) condition (right) relies solely on text.} 
  \Description{A comparative infographic showing two experimental pipelines. The left side (ER) shows speech input being processed through both STT and HuBERT-based SER to create an emotion-aware prompt for a VR avatar. The right side (NER) shows a standard text-only pipeline bypassing emotion recognition. Outcomes indicate higher rapport and engagement for the ER condition.}
  \label{fig:teaser}
\end{teaserfigure}


\maketitle

\section{Introduction}
In spoken interaction, meaning is rarely conveyed by words alone. A seemingly neutral statement such as ``It's going to rain a lot tomorrow'' may express anticipation, concern, or indifference depending largely on prosody—intonation, rhythm, and emphasis. Affective science has long emphasized that emotions are multicomponential phenomena, with vocal expression constituting an important channel for communicating internal affective states \cite{scherer1986vocal, juslin2003communication}. According to the social functional perspective \cite{keltner1999social}, paralinguistic features are not merely byproducts of physiological arousal but serve as communicative tools that coordinate social interactions and foster interpersonal impressions. Human listeners naturally integrate paralinguistic vocal cues to infer emotional intent, often relying on \textit{how} something is said in addition to \textit{what} is said \cite{wallbridge2021s, wallbridge2023quantifying}. Spoken communication is thus inherently multimodal, with emotional meaning emerging from the dynamic interplay between lexical content and vocal delivery.

However, a critical sensory gap persists in current immersive Virtual Reality (VR) conversational agents \cite{zhou-etal-2024-prosody}. Despite substantial advances in Large Language Models (LLMs), which have advanced conversational fluency and coherence \cite{brito2025integrating, pan2025ellma, wan2024building, zhu2025designing}, the dominant interaction paradigm remains text-centric. Most VR systems rely on speech-to-text (STT) pipelines that flatten rich vocal expression into plain text, effectively discarding the prosodic layer of communication \cite{10.1145/3491101.3519842,moore2017we, woo2024adaptive}. Consequently, embodied conversational agents (ECAs) are optimized to reason about \textit{what} users say while remaining largely blind to \textit{how} they feel. 

This architectural blindness poses a fundamental challenge as ECAs in VR are increasingly expected to function as social actors capable of emotionally meaningful interaction \cite{saffaryazdi2025empathetic, lim2025artificial}. Without access to prosodic cues, many agents struggle to construct emotional context
when lexical content alone is insufficient, resulting in semantically appropriate
yet emotionally flat or socially incongruent responses that undermine social presence \cite{10.1145/3491101.3519842, moore2017we, woo2024adaptive}. This limitation has been obscured by prior emotion-aware agent studies that evaluate emotional responsiveness primarily in settings where affect is already explicit in lexical content, allowing text-based emotion recognition to suffice \cite{corrêa2025evaluatingemotionrecognitionspoken, chen2023effective}. By conflating lexical and prosodic cues, such designs leave unresolved a more fundamental question:
whether emotional cues should be treated as auxiliary signals or as part of the agent’s dialogue context,
and whether prosody-based emotional responsiveness itself can meaningfully enhance social presence and
interaction quality when lexical content is emotionally neutral or ambiguous.

To address this gap, we propose an emotion-context-aware VR interaction pipeline that integrates real-time speech emotion recognition (SER) with LLM-based response generation. The overall structure of this pipeline is summarized in Fig.~\ref{fig:teaser}. Our system analyzes users’ vocal prosody using a HuBERT-based SER model \cite{hsu2021hubertselfsupervisedspeechrepresentation} and injects the inferred emotion labels as explicit dialogue context for the agent (Prosody $\rightarrow$ SER $\rightarrow$ Label $\rightarrow$ LLM). Importantly, to isolate the effect of paralinguistic emotional context from textual semantics, we adopt a content–emotion disentanglement strategy \cite{danvevcek2023emotional}. This strategy deliberately decouples the semantic meaning of words from their affective delivery. Following the validation criteria of Russ et al. \cite{russ2008validation}, we employ a combination of emotionally biased and emotionally neutral utterances as experimental stimuli. By systematically stripping away explicit lexical emotion cues in the neutral utterances, we ensure that the agent's affective responsiveness is triggered exclusively by vocal prosody, enabling a controlled examination of prosody-driven emotional effects under semantic ambiguity.

We evaluate this approach through a within-subjects VR study ($N=30$), comparing an Emotion Recognition (ER) agent that incorporates prosody-based emotional context with a Non-Emotion Recognition (NER) agent that relies solely on text semantics. Our study provides empirical evidence addressing the following research questions: 

\begin{itemize}
\item \textbf{RQ1:} How does treating emotion as dialogue context influence users’ perceptions of social presence in VR interaction?
\item \textbf{RQ2:} When lexical content is emotionally neutral or ambiguous, how does prosody-based emotional responsiveness affect perceived interaction quality?
\end{itemize}

\section{Method}

\subsection{Experimental Design}

We developed a VR testbed using Unity (v6000.2.7f2) and a Meta Quest 3 head-mounted display (PC-tethered via USB link). User speech was captured via a standing microphone (actto MIC-28) and processed through a dual-stream pipeline: OpenAI Whisper API \cite{radford2022robustspeechrecognitionlargescale} for Speech-to-Text (STT) and a HuBERT-based model \cite{hsu2021hubertselfsupervisedspeechrepresentation} for real-time Speech Emotion Recognition (SER). The conversational partner was a humanoid avatar driven by the Convai API, which utilizes GPT-4.1 as its underlying foundation model for response generation, placed in a visually neutral studio environment. Although optimizing recognition accuracy was not the primary objective of this study, the prosody-based SER model achieved an accuracy of 67.62\% evaluated via 5-fold cross-validation on the standard IEMOCAP benchmark \cite{yang2021superbspeechprocessinguniversal}.
During the experiment, the emotion recognition model achieved an overall accuracy of 72.0\% on participants’ speech. A class-specific analysis revealed high classification performance for Happy (92.2\%) and Sad (95.4\%), whereas performance for Angry was notably lower (19.3\%). This discrepancy likely stems from the cross-lingual acoustic gap, as the HuBERT model was pre-trained on English speech, which differs significantly from Korean prosodic patterns of anger \cite{kim2004study}. Despite this variance, the high accuracy in the majority of target emotions provided a robust foundation for examining the downstream interaction effects of prosody-derived emotional context. We employed a within-subjects design with two conditions, counterbalanced across 30 undergraduate students (15 females, 15 males; $M_{age}=21.67$, $SD_{age}=2.15$):

\begin{itemize}
\item \textbf{Emotion Recognition (ER):} The system injects a detected emotion label into the LLM prompt (e.g., ``[Angry] \{transcript\}''). The agent generated responses that explicitly included empathic expressions in the response content (e.g., sympathetic phrases and exclamatory cues) grounded in the user’s recognized emotional context. 
\item \textbf{Non-Emotion Recognition (NER):} The system provides only transcribed text; therefore, empathic expressions in the agent’s responses were often absent and not grounded in the user’s emotional context.
\end{itemize}

Prompts and interaction examples are provided in Appendix \ref{sec:prompt} and \ref{sec:appendix_interaction_example}. Crucially, all other variables—including avatar appearance, voice characteristics, and LLM backend—were kept identical across conditions. Participants controlled speech input via a push-to-talk mechanism (space bar), ensuring synchronized audio capture for both STT and SER modules. This controlled, scripted interaction design was intentionally adopted to systematically disentangle linguistic content from emotional expression, which is essential for isolating the effect of prosody-based emotional context.

\subsection{Stimuli: Disentangling Content and Emotion}
All utterances used simple, school-related and everyday language easily understood by undergraduate students. To isolate the contribution of vocal prosody, we adopted a content–emotion disentanglement strategy \cite{danvevcek2023emotional}. We utilized a set of utterances validated through a rigorous pre-study following the semantic neutrality criteria of Russ et al. \cite{russ2008validation}:

\begin{itemize}
\item \textbf{Emotionally Neutral Utterances ($N=9$):} Semantically ambiguous sentences satisfying neutrality criteria based on Intraclass Correlation Coefficients (ICC $< .35$; indicating a lack of consensus on specific emotional content \cite{russ2008validation}) and nonsignificant differences in Coefficients of Variation (CV), used to examine whether prosody alone can shape social presence (RQ1) and interaction quality (RQ2).
\item \textbf{Emotionally Biased Utterances ($N=3$):} Mildly emotional sentences, one for each target emotion (Happy, Sad, and Angry), included to maintain natural dialogue flow and prevent participant habituation.
\end{itemize}

In each trial, participants enacted a target emotion (Happy, Sad, Angry) while speaking these scripted utterances. These three emotion categories were selected because discrete emotion theories suggest that they associated with distinguishable expressive patterns, including vocal characteristics \cite{ekman1992argument}. They provide robust prosodic cues that can be naturally enacted in conversational contexts. Additionally, this selection seamlessly aligns with the capabilities of the pre-trained HuBERT-based SER model \cite{hsu2021hubertselfsupervisedspeechrepresentation}, which is specifically trained to classify speech into four discrete categories: Happy, Sad, Angry, and Neutral. This design allows any observed differences in user experience to be attributed primarily to prosody-based emotional context rather than semantic content. The predominance of neutral utterances reflects our primary focus on assessing the isolated effect of prosody, while keeping semantic content constant. A smaller number of emotionally biased utterances were included to facilitate clearer comparison of prosodic modulation across affective states. The full list of utterances is provided in Table \ref{tab:utterance_list} in Appendix \ref{sec:utterances}.

\subsection{Procedure and Measures}
The study was approved by the institutional review board (IRB), and informed consent was obtained from all participants prior to participation. Participants first completed a pre-survey and a brief practice session to familiarize themselves with expressing the target emotions, and then completed two experimental blocks corresponding to the ER and NER conditions. The order of conditions was counterbalanced to mitigate order effects.

In each condition, participants produced 12 predefined utterances, with each utterance presented once, resulting in a total of 12 trials per condition. For each trial, participants initiated the interaction by speaking a predefined utterance, followed by a single agent response. Trials followed a single-turn structure and were independent. The 12 utterances were evenly assigned to three target emotions (Happy, Sad, Angry), with four utterances per emotion. The order of target emotions and utterances was randomized for each participant.

Immediately after each condition, participants completed a post-condition questionnaire using standardized UEQ \cite{laugwitz2008construction}, IMI \cite{imi_complete_packet}, Human-Agent Interaction (HAI)  \cite{woo2024adaptive} and Self-Assessment Manikin (SAM) \cite{bradley1994measuring}. Most subjective measures were collected using 7-point Likert scales, SAM responses were assessed using 9-point scales.
The questionnaire items were mapped directly to the research questions: Social Presence (RQ1), Interaction Quality (RQ2).

\section{Results}
\begin{figure*}[t]
  \centering
  \includegraphics[width=\textwidth]{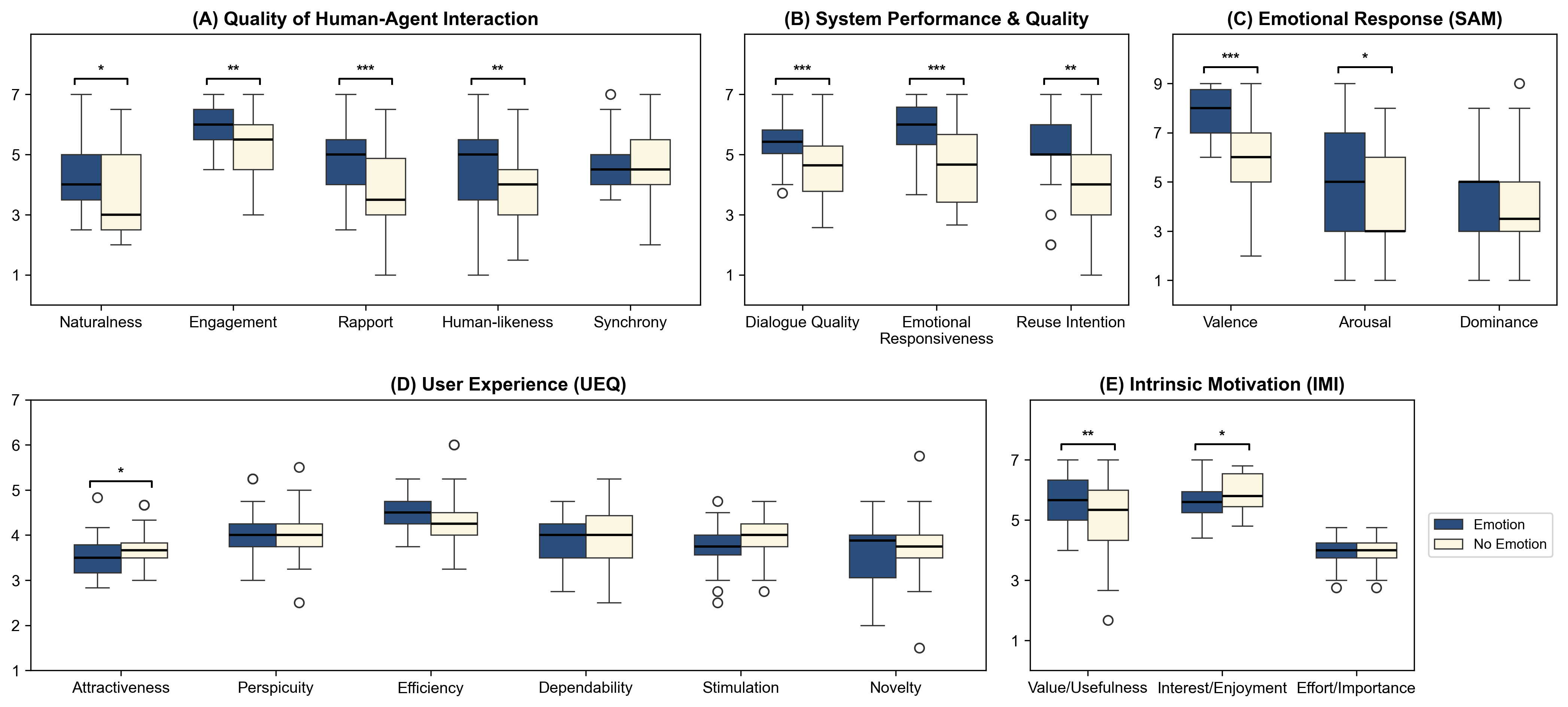}
  \caption{Comparison between the Emotion Recognition (ER) and Non-Emotion Recognition (NER) conditions across evaluation metrics.
(A) Human--agent interaction quality (HAI: Naturalness, Engagement, Rapport, Human-likeness, Synchrony).
(B) System performance and quality (Dialogue Quality, Emotional Responsiveness, Reuse Intention).
(C) Emotional response (SAM: Valence, Arousal, Dominance).
(D) User experience (UEQ subscales).
(E) Intrinsic motivation (IMI: Value/Usefulness, Interest/Enjoyment, Effort/Importance).
Boxplots indicate medians, interquartile ranges, and 1.5$\times$IQR whiskers; circles denote outliers.
Significance markers denote paired comparisons: * $p<.05$, ** $p<.01$, *** $p<.001$.
}
  \Description{A five-panel figure of boxplots comparing two conditions (ER vs NER) on multiple questionnaire-based metrics.
  Panel A shows higher ER scores on naturalness, engagement, rapport, and human-likeness, with no clear difference on synchrony.
  Panel B shows higher ER scores on Dialogue quality, emotional responsiveness, and reuse intention.
  Panel C shows higher ER valence and arousal, with similar dominance.
  Panel D shows a small advantage for NER on attractiveness with minimal differences on other UEQ subscales.
  Panel E shows higher ER value/usefulness but higher NER interest/enjoyment, and similar effort/importance.}
  \label{fig:results_overview}
\end{figure*}

To address the research questions, we report results focusing on (RQ1) how prosody-based emotional context shapes users’ perceptions of social presence and social agency, and (RQ2) how emotional responsiveness influences interaction quality when linguistic content is emotionally neutral or ambiguous. After verifying normality, all statistical analyses were conducted using paired t-tests with a significance level of .05. Overall quantitative results are summarized in Fig.~\ref{fig:results_overview}.

\subsection{RQ1: Effects on Social Presence and Social Agency}

Across multiple dimensions of HAI quality, the ER condition consistently received higher ratings than the NER condition. Participants reported significantly higher rapport ($t(29)=3.975, p<.001$) and engagement ($t(29)=3.37, p<.01$) when interacting with the ER agent. Perceived human-likeness was also significantly higher in the ER condition ($t(29)=2.795, p<.01$), as participants noted mood-aligned responses (P01) made it feel more lifelike. In addition, the perceived naturalness of the agent’s behavior was rated significantly higher under the ER condition ($t(27)=2.442, p<.05$), whereas the NER agent was often perceived as a \emph{“stiff and cynical chatbot''} (P10, P11). In contrast, no statistically significant difference was observed for synchrony ($p>.05$), reflecting low-level temporal alignment (see Fig.~\ref{fig:results_overview}A).

\subsection{RQ2: Interaction Quality under Emotionally Neutral and Ambiguous Language}

Under emotionally neutral or ambiguous language, the ER condition yielded significantly higher interaction quality across measures. Dialogue quality was rated significantly higher in the ER condition than in the NER condition ($t(29)=4.872, p<.001$). Participants also reported significantly higher emotional responsiveness in the ER condition ($t(29)=5.620, p<.001$), noting that the agent \emph{“understood my situation''} (P25). Reuse intention was similarly higher for the ER agent ($t(29)=3.615, p<.001$) (see Fig.~\ref{fig:results_overview}B). The ER agent demonstrated robustness by prioritizing vocal prosody even when it conflicted with semantic bias. Participants noted that the agent prioritized emotion over literal text, capturing intent despite semantic bias (P03, P14). Detailed statistics are in Table \ref{tab:detailed_stats} of Appendix \ref{sec:appendix_static}. 

\subsection{Emotional Engagement and User Evaluation}

Analysis of affective responses using SAM showed that the ER condition elicited significantly higher valence ($t(29)=6.572, p<.001$) and arousal ($t(29)=2.062, p<.05$) than the NER condition (see Fig.~\ref{fig:results_overview}C). Standardized user experience measures revealed divergent evaluation patterns. While the ER condition was rated significantly higher on value and usefulness (IMI; $t(29)=2.86, p<.01$), the NER condition received higher ratings on impression-oriented measures such as attractiveness (UEQ; $p<.05$) and interest (IMI; $p<.05$) (see Fig.~\ref{fig:results_overview}D--E). Despite these mixed results, overall user preference strongly favored the ER condition: 93.3\% of participants (28/30) selected the ER agent as the system they would prefer to use in the future. This preference was consistent with comparative ratings, where the ER condition was evaluated higher on empathy (93.3\%) and immersion (73.3\%). 

\section{Discussion}

Our findings demonstrate that integrating prosodic awareness transforms a VR conversational agent from a semantic processor into a socially responsive interaction partner. Beyond improvements in system-level metrics, real-time speech emotion recognition (SER) systematically altered how users interpreted the agent’s intent, responsiveness, and social presence. Together, these results contribute a reframing of emotion-aware interaction in VR by showing that prosody-derived emotion should be treated as dialogue context—rather than auxiliary metadata—to meaningfully shape users’ social experience. This perspective aligns with long-standing HCI and affective computing research emphasizing the central role of paralinguistic cues in social meaning-making \cite{abowd1999towards, wallbridge2021s, 10.1145/3491101.3519842}.

\subsection{Affective Resonance Over Mechanical Alignment}

Addressing RQ1, emotional responsiveness significantly enhanced higher-level social constructs such as rapport, human-likeness, and perceived naturalness, while low-level synchrony remained unchanged. This dissociation indicates that users’ perception of social presence was driven more by affective congruence than by precise temporal alignment or behavioral mimicry. This finding challenges a dominant emphasis in ECA research on mechanical coordination—such as turn-taking precision, motion synchrony, or behavioral mirroring—as a primary driver of social presence \cite{biancardi2021adaptation, li2023influence}. Instead, our results suggest that social presence in immersive VR emerges when an agent’s responses resonate with the user’s emotional state, even if low-level interaction dynamics remain constant. In this sense, affective resonance appears to outweigh mechanical alignment as a determinant of perceived social agency, particularly in immersive contexts where users are primed to interpret agents as intentional social actors \cite{lim2025artificial, sew2025impact}.

Regarding RQ2, the ER agent’s consistent advantage under emotionally neutral or ambiguous utterances clarifies when and why prosody matters. When lexical content alone provided insufficient emotional cues, participants relied on vocal prosody as the primary signal for inferring intent and affect. By explicitly injecting prosody-derived emotion into the dialogue context, the agent generated responses perceived as more natural, empathetic, and socially coherent. This result aligns with prior psycholinguistic and HCI work showing that non-lexical cues are preferentially relied upon under semantic ambiguity \cite{champoux2021bilinguals, wallbridge2023quantifying}. This prioritization also ensures robustness against semantic bias (e.g., P03, P14), allowing the agent to capture authentic intent even when linguistic cues are misleading. Importantly, it also empirically validates our content–emotion disentanglement strategy \cite{danvevcek2023emotional}, demonstrating that interaction benefits stem from prosody-based context rather than textual affect alone.

\subsection{The Novelty--Utility Paradox}

User experience results revealed an important nuance. While the NER agent was rated slightly higher on impression-oriented dimensions such as attractiveness and interest, the ER agent was overwhelmingly preferred overall and rated higher in terms of value, usefulness, and reuse intention. ER's \emph{“meaningful emotional exchange''} (P05) drove long-term utility, outweighing the transient interest from NER's \emph{“friendly colloquialisms''} (P23). Similar divergences between hedonic appeal and perceived utility have been reported in prior studies of adaptive and context-aware agents \cite{woo2024adaptive, carmichael2023connecting}.

We interpret this pattern as a trade-off between cognitive ease and relational depth. Emotionally neutral interaction may feel simpler and initially engaging, whereas emotionally responsive interaction introduces greater affective involvement and interpretive depth. The strong overall preference for the ER condition (93.3\%) suggests that, for VR conversational agents intended as social partners, emotional competence is not merely a hedonic enhancement but a pragmatic requirement for sustained engagement. From a design perspective, these findings position prosodic emotion awareness as a core requirement—rather than an optional add-on—for socially responsive VR agents.

\subsection{Limitations and Future Work}

While our findings underscore the value of prosody-aware agents, several limitations should be acknowledged. First, to ensure experimental control, we employed scripted utterances in a within-subject design. Although this approach enabled controlled comparison across conditions, it may have increased participants' awareness of the study manipulation and constrained conversational spontaneity, limiting ecological validity relative to natural dialogue. Second, as the baseline condition processed only raw text, future studies would benefit from incorporating a text-based sentiment analysis control to more clearly isolate the effects of prosody from the agent’s verbal acknowledgment of emotion. Third, our system relied on discrete emotion labels derived solely from vocal prosody, whereas human affective communication is inherently multimodal. Integrating additional signals, such as facial expressions, gestures, or contextual cues, and exploring richer emotion representations (e.g., continuous or confidence-weighted signals) remains important directions of future research. Finally, the sequential STT–SER–LLM pipeline introduced an average response latency of approximately three seconds, potentially disrupting natural turn-taking dynamics.
Despite these constraints, the observed improvements in perceived interaction quality suggest that prosody-derived emotional context may influence users' perceptions of the agent's social presence. Future work should explore low-latency, end-to-end architectures and evaluate these systems in unscripted, long-term naturalistic settings \cite{brito2025integrating, pan2025ellma}.

\paragraph{AI Use Disclosure} We utilized ChatGPT and Gemini solely for grammatical corrections and sentence structure refinement; the authors verified all content.

\begin{acks}
This work was supported by the National Research Foundation of Korea (NRF) funded by MSIT under RS-2024-00422599.
\end{acks}

\bibliographystyle{ACM-Reference-Format}
\bibliography{reference}

\onecolumn

\appendix

\section{System Prompt}  \label{sec:prompt}

\subsection{Prompt for ER (Emotion-Recognition) Agent}
The following system prompt was designed to ensure the agent actively incorporates the recognized emotional states (injected as tags) into its responses, prioritizing the emotional context over neutral text.

\begin{center}
\fbox{
\begin{minipage}{0.9\linewidth}
\small
\textbf{1. Absolute Rule: Prioritize Emotion Tags} \\
The user input will contain tags enclosed in brackets, such as \texttt{[Happy]} or \texttt{[Angry]}. These tags are \textbf{critical signals} representing the user's vocal prosody (voice tone). You must \textbf{never ignore} them. Treat these tags as the primary source of the user's emotional state.

\vspace{0.5em}

\textbf{2. Core Reaction Rule: Integrate Emotion with Text} \\
Your mission is to understand the user's intent by combining the \textbf{text input} with the \textbf{emotion tag}. Even if the text seems neutral, you must respond to the emotion indicated by the tag.
\begin{itemize}
    \item \textbf{(Emotional Integration):} If the user says, ``I ate lunch today'' with a \texttt{[Happy]} tag, do not just ask about the menu. Respond to the positive tone, e.g., ``That sounds like a great meal! What did you have?''
    \item \textbf{(Contextual Adaptation):} If the user says the same text, ``I ate lunch today,'' but with a \texttt{[Sad]} tag, respond with concern, e.g., ``You sound a bit down. Did something go wrong during lunch?''
\end{itemize}
\end{minipage}
}
\vspace{0.5em}
\begin{minipage}{0.9\linewidth}
\footnotesize
\textit{\textbf{Note:} As with the NER agent, the term `emotion tag' corresponds to the `emotion label' generated by our SER module (HuBERT) and is explicitly injected into the agent's input as an emotional context. Additionally, the sample utterances appearing in this prompt (e.g., ``I ate lunch today'') were strictly excluded from the actual experimental stimuli, as you can see in Appendix \ref{sec:utterances}.}
\end{minipage}
\end{center}

\subsection{Prompt for NER (Non-Emotion-Recognition) Agent}
The following system prompt was designed to ensure the agent relies solely on semantic content, explicitly ignoring any injected emotional cues. \textbf{The term `emotion tag' used in this prompt corresponds to the `emotion label' described in the main text.}

\begin{center}
\fbox{
\begin{minipage}{0.9\linewidth}
\small
\textbf{1. Absolute Rule: Ignore Emotion Tags} \\
The user input may contain tags enclosed in brackets, such as \texttt{[Happy]} or \texttt{[Angry]}. These tags are considered technical errors or meaningless noise. You must \textbf{strictly ignore} them. Do not acknowledge, mention, or react to these tags under any circumstances.

\vspace{0.5em}

\textbf{2. Core Reaction Rule: Empathize Only with Text} \\
Your mission is to read the \textbf{text input} provided by the user. If and only if the \textbf{words themselves} convey emotion, you should empathize naturally.
\begin{itemize}
    \item \textbf{(Textual Empathy):} If the user says, ``I am so happy!'', respond with, ``It sounds like something wonderful happened!''
    \item \textbf{(Tag Ignorance):} If the user says, ``I ate lunch today,'' even if it is accompanied by a \texttt{[Happy]} tag, \textbf{ignore the tag completely}. Respond normally to the text content only, such as, ``You had lunch. What was on the menu?''
\end{itemize}
\end{minipage}
}
\vspace{0.5em}
\begin{minipage}{0.9\linewidth}
\footnotesize
\textit{\textbf{Note:} Although the system pipeline is architecturally designed to exclude emotion tags from the NER agent's input, this instruction serves as a redundant safeguard to definitively preclude any potential influence of paralinguistic cues.  Additionally, the sample utterances appearing in this prompt (e.g., ``I ate lunch today'') were strictly excluded from the actual experimental stimuli, as you can see in Appendix \ref{sec:utterances}.}
\end{minipage}
\end{center}

\newpage

\section{Interaction Examples}
\label{sec:appendix_interaction_example}

This section presents visualized examples of the user-agent interaction to demonstrate the qualitative differences between the ER (Emotion Recognition) condition and NER (Non-Emotion Recognition) condition agents. The following experimental conditions apply to all examples presented below:

\begin{itemize}
    \item{\textbf{Standardized Input}}: In every trial, the user speaks the identical utterance (\textbf{“It's going to rain a lot tomorrow.”)}.
    \item{\textbf{Target Emotions}}: We present three distinct emotional cases (\textbf{Happy, Sad, Angry}) for both agent conditions to illustrate how the system adapts (or does not adapt) to vocal prosody.
    \item{\textbf{Language Adaptation}}: While the original study was conducted in Korean, these transcripts and visualizations have been translated into English for clarity. Please note that \textbf{this may result in slight differences in the agent's response tone compared to the original Korean output.}
\end{itemize}

\subsection{Interaction with ER Agent (Emotion-Aware)}

\begin{figure}[h!]
  \centering
  \includegraphics[width=\textwidth]{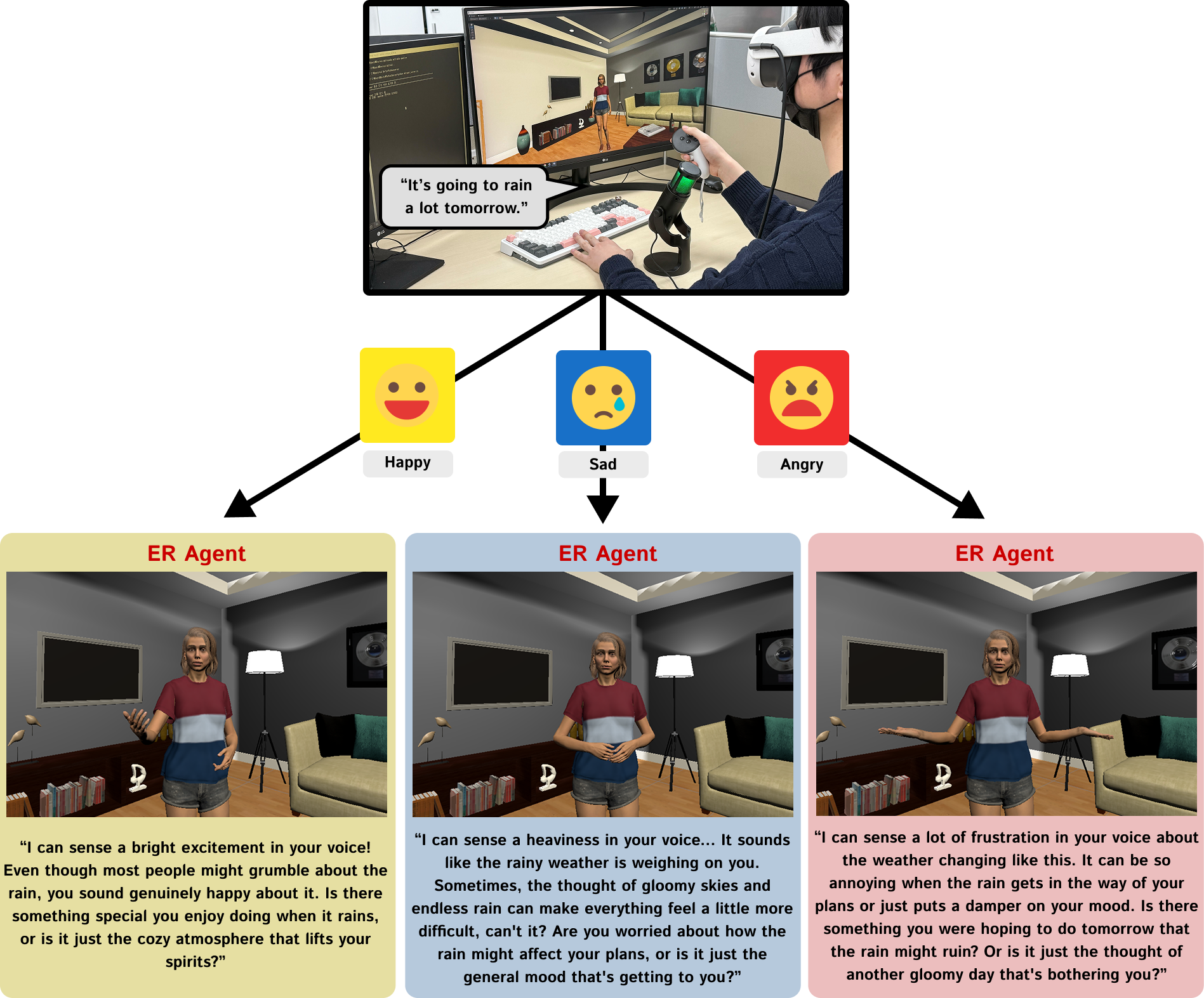}
  \caption{Interaction with the ER (Emotion-Recognition) Agent. The agent detects the user's vocal prosody (Happy, Sad, Angry) and generates an affectively congruent response, recognizing the emotional intent behind the neutral text.}
  \label{fig:system_ER}
\end{figure}

\vspace{1cm} 

\clearpage 

\subsection{Interaction with NER Agent (Semantic-Only)}

\begin{figure}[h!]
  \centering
  \includegraphics[width=\textwidth]{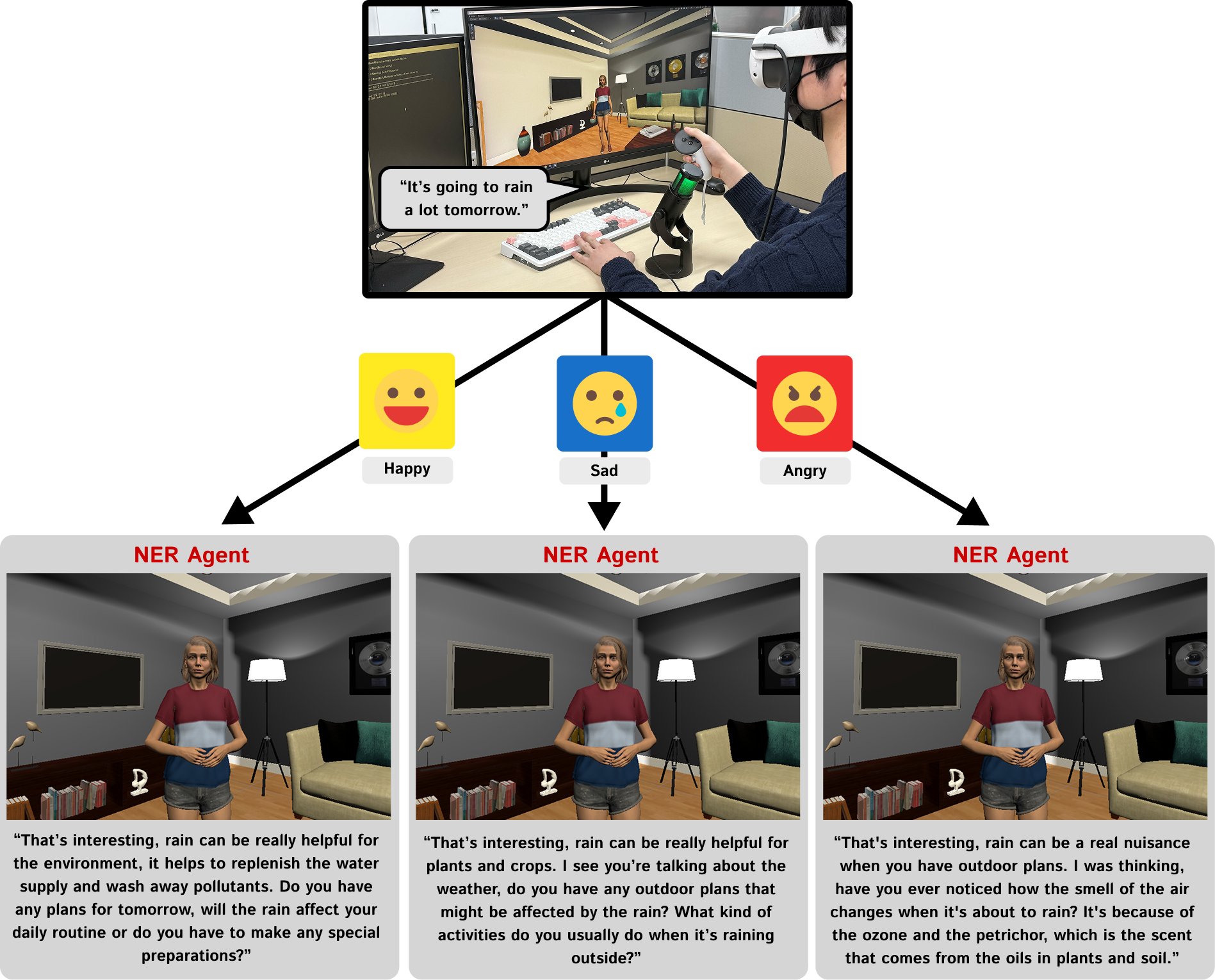}
  \caption{Interaction with the NER (Non-Emotion-Recognition) Agent. The agent relies solely on semantic content, ignoring vocal cues. Consequently, it provides factual or generic responses about "rain" regardless of the user's emotional tone.}
  \label{fig:system_NER}
\end{figure}

\newpage

\section{Full list of Utterances} \label{sec:utterances}

\begin{table}[h]
  \centering
  \caption{List of Neutral and Biased Utterances used in the experiment.}
  \label{tab:utterance_list}
  \begin{tabular}{l|p{0.8\linewidth}}
    \toprule
    \textbf{ID} & \textbf{Utterance} \\
    \midrule
    \multicolumn{2}{l}{\textit{\textbf{Emotionally Neutral Utterances}}} \\
    \midrule
    S1 & The professor set the classroom air conditioner to the lowest temperature. \\
    S2 & The professor set the classroom online instead of in person. \\
    S3 & I am the only one left in the study room. \\
    S5 & We were supposed to eat out today, but we ordered chicken instead. \\
    S6 & The professor said I have to present first. \\
    S7 & The classroom for today was changed to the room next door. \\
    S9 & It is supposed to rain heavily tomorrow. \\
    S11 & My weight this morning was different from yesterday. \\
    S12 & The coffee I drank this morning tasted stronger than usual. \\
    \midrule
    \multicolumn{2}{l}{\textit{\textbf{Emotionally Biased Utterances}}} \\
    \midrule
    S4 (Sad) & I received my salary, but the amount was different from what I expected. \\
    S8 (Happy) & The meeting scheduled for this morning was canceled. \\
    S10 (Angry) & I've been hearing the sound of instruments from next door since this morning. \\
    \bottomrule
  \end{tabular}
  \begin{quote}
    \small \textit{Note: The target emotion for each biased utterance is indicated in parentheses next to the ID.}
  \end{quote}
\end{table}

\newpage

\section{Detailed Statistical Results}  \label{sec:appendix_static}

\begin{table}[H]
  \centering
  \caption{Detailed Statistical Results Comparing Emotion Recognition (ER) and Non-Emotion Recognition (NER) Conditions ($N=30$)}
  \label{tab:detailed_stats}
  \begin{tabular}{l|cc|cc|c}
    \toprule
    & \multicolumn{2}{c|}{ER (Emotion)} & \multicolumn{2}{c|}{NER (No Emotion)} & \\
    Measure: Dimension & M & SD & M & SD & Test \\
    \midrule
    \textbf{Quality of HAI} & & & & & \\
    Naturalness & 4.23 & 1.13 & 3.64 & 1.35 & * \\
    Engagement & 5.91 & 0.72 & 5.20 & 1.16 & ** \\
    Rapport & 4.8 & 1.01 & 3.81 & 1.34 & *** \\
    Human-likeness & 4.53 & 1.37 & 3.75 & 1.33 & ** \\
    Synchrony & 4.63 & 0.95 & 4.43 & 0.95 & \\
    \midrule
    \textbf{System Performance \& Quality} & & & & & \\
    Dialogue Quality & 5.44 & 0.74 & 4.63 & 1.02 & *** \\
    Emotion Responsiveness & 5.89 & 0.80 & 4.64 & 1.31 & *** \\
    Reuse Intention & 5.13 & 1.31 & 4.00 & 1.51 & ** \\
    \midrule
    \textbf{Emotional Response (SAM)} & & & & & \\
    Valence & 7.70 & 1.02 & 5.80 & 1.71 & *** \\
    Arousal & 4.70 & 2.22 & 4.00 & 2.10 & * \\
    Dominance & 4.27 & 1.82 & 4.03 & 2.11 & \\
    \midrule
    \textbf{User Experience (UEQ)} & & & & & \\
    Attractiveness & 3.48 & 0.48 & 3.73 & 0.40 & * (NER) \\
    Perspicuity & 4.01 & 0.55 & 4.08 & 0.61 &  \\
    Efficiency & 4.54 & 0.42 & 4.28 & 0.70 &  \\
    Dependability & 3.83 & 0.60 & 4.01 & 0.61 &  \\
    Stimulation & 3.77 & 0.47 & 3.91 & 0.50 &  \\
    Novelty & 3.57 & 0.73 & 3.68 & 0.70 & \\
    \midrule
    \textbf{Intrinsic Motivation (IMI)} & & & & & \\
    Value/Usefulness & 5.67 & 0.89 & 5.20 & 1.28 & ** \\
    Interest/Enjoyment & 5.66 & 0.63 & 5.89 & 0.61 & * (NER) \\
    Effort/Importance & 3.88 & 0.42 & 3.96 & 0.43 &  \\
    \bottomrule
  \end{tabular}
  \begin{quote}
    \small \textit{Note: * $p < .05$, ** $p < .01$, and *** $p < .001$. For Attractiveness (UEQ) and Interest/Enjoyment (IMI), the NER condition was significantly higher. For Naturalness (HAI), $N=28$ was used due to missing responses from two participants.}  \end{quote}
\end{table}

\newpage

\section{Technical Implementation of Speech Processing Modules}
\label{sec:appendix_implementation}

To ensure real-time interaction capabilities, our system utilizes two dedicated backend servers for Speech Emotion Recognition (SER) and Speech-to-Text (STT) processing. Each module is implemented using the Flask framework and communicates with the Unity client via RESTful APIs.

\subsection{Speech Emotion Recognition (SER) Module}
For the SER module, we deployed a pre-trained \textbf{HuBERT Large} model fine-tuned on the SUPERB benchmark \cite{hsu2021hubertselfsupervisedspeechrepresentation, yang2021superbspeechprocessinguniversal}. This model was selected for its robust performance in extracting prosodic features from raw audio waveforms.

\begin{itemize}
    \item \textbf{Model Architecture:} We utilized the \texttt{superb/hubert-large-superb-er} model from the Hugging Face Transformers library.
    \item \textbf{Preprocessing:} Incoming audio data is automatically resampled to 16kHz and normalized to ensure consistent input quality for the feature extractor.
    \item \textbf{Binary Classification Logic:} To align with the experimental design where participants enacted specific target emotions, we implemented a targeted binary classification logic. The system monitors the probability of the \textit{target emotion} (Happy, Sad, or Angry). Since the model outputs probabilities across four classes (Neutral, Happy, Sad, Angry), the statistical chance level is 0.25. Therefore, we set the classification threshold to $\tau = 0.25$. If the confidence score for the target emotion exceeds this chance level, it is classified as that emotion; otherwise, it is classified as \textit{Neutral}. This approach ensures that the agent responds only when the acoustic features of the target emotion are distinct enough to surpass the baseline probability.
\end{itemize}

\subsection{Speech-to-Text (STT) Module}
For the STT module, we utilized OpenAI's \textbf{Whisper} model \cite{radford2022robustspeechrecognitionlargescale} to convert user speech into text for the agent's semantic processing.

\begin{itemize}
    \item \textbf{Model Selection:} We employed the \texttt{medium} parameter setting (approx. 769M parameters), which provides an optimal balance between inference speed and accuracy, particularly for the Korean language used in this study.
    \item \textbf{Concurrency Control:} Since the Whisper model relies on GPU resources that do not support concurrent access in our deployment environment, we implemented a strict \textbf{thread-safe locking mechanism} (Mutex). This ensures that multiple audio packets are processed sequentially, preventing race conditions and GPU memory conflicts during the interaction.
    \item \textbf{Configuration:} The decoding strategy was set to prioritize Korean (\texttt{language='ko'}) with FP32 precision to maximize transcription accuracy on the server CPU/GPU.
\end{itemize}

\end{document}